\documentstyle[12pt]{article}
\newenvironment{eq}
{\[\begin{array}}{\end{array}\]{}}

\let\rvec=\vec        % save plain's \vec as \rvec
   \def\({\Bigl(}
\def\){\Bigr)}    \def\|{\Big|}
\def\then{~\Rightarrow~}   \def\o{\circ}
\def\m{\bullet}    \def\x{\times}
   \def\ox{\otimes}
 
\def\pl{{~\oplus~}}

\def\mid{\big\bracevert}

\def\subnoteq{\subset}

\def\and{\wedge}

\def\AND{\displaystyle\bigwedge}
\def\od{\vee}

\def\OD{\displaystyle\bigvee}

\def\rin{{\,\in\kern-.42em\in}}

\def\tr{{\,{\rm tr }\,}}
\def\det{\,{\rm det }\,}
\def\id{\,{\rm id}}

\def\sx{~\rvec\x~\!}
\def\weights{{{\bf weights}\,}}

\def\A{{\,{\rm A\kern-.55emA}}}
\def\B{{\,{\rm I\kern-.2emB}}}
\def\C{{\,{\rm I\kern-.55emC}}}
\def\E{{\,{\rm I\kern-.2emE}}}
\def\G{{\,{\rm I\kern-.55emG}}}
\def\H{{{\rm I\kern-.2emH}}}
\def\I{{\,{\rm I\kern-.2emI}}}
\def\K{{\,{\rm I\kern-.2emK}}}
\def\L{{\,{\rm I\kern-.2emL}}}
\def\M{{\,{\rm I\kern-.16emM}}}
\def\N{{\,{\rm I\kern-.16emN}}}
\def\Q{{\,{\rm I\kern-.5emQ}}}
\def\R{{{\rm I\kern-.2emR}}}
\def\S{{\,{\rm I\kern-.42emS}}}
\def\T{{\,{\rm I\kern-.37emT}}}
\def\UU{{\,{\rm I\kern-.51emU}}}
\def\Z{{\,{\rm Z\kern-.32emZ}}}

\def\al{\alpha}  \def\be{\beta} \def\ga{\gamma}
\def\de{\delta}  \def\ep{\epsilon}  \def\ze{\zeta}
    \def\io{\iota}
   \def\la{\lambda}   \def\si{\sigma}
   \def\om{\omega} 
\def\phi{\varphi}  \def\Ga{\Gamma}  
    \def\La{\Lambda}

\def\vec#1{\underline{\bf vec}_{#1}}

\def\GL{{\bf GL}}  
\def\SL{{\bf SL}}
\def\U{{\bf U}} 
 
\def\O{{\bf O}}   
\def\SU{{\bf SU}} 
 
\def\SO{{\bf SO}}

\def\AL{{\bf AL}}

\def\d#1{{\check{#1}}}

\def\ty#1{{\tt #1}}
\def\ro#1{{\rm #1}}
\def\bl#1{{\bf {#1}}}
\def\cl#1{{\cal #1}}

\def\ol#1{\overline{#1}}

\def\dprod#1#2{\langle#1,#2\rangle}
\def\sprod#1#2{\langle#1|#2\rangle}

\def\com#1#2{\lbrack#1,#2\rbrack}

\def\map{\longrightarrow}
\def\inmap{\hookrightarrow}
\def\lrmap{\leftrightarrow}
\def\llrmap{\lmap\hskip-4mm\map}
\def\lmap{\longleftarrow}
\def\dmap{\Big\downarrow}

\def\mape{\longmapsto}

\def\Diagr#1#2#3#4#5#6#7#8{\matrix{\noalign{\vskip5mm}
      &              &{\scriptstyle #5}&              &     \cr
      & #1           & \map           & #2           &     \cr
{\scriptstyle #8}   &\dmap         &    &\dmap  &{\scriptstyle#6} \cr
      & #4           & \map           & #3           &     \cr
      &              &{\scriptstyle#7}&              &     \cr
\noalign{\vskip5mm}             }}

\def\Diagrlrud#1#2#3#4#5#6#7#8{\matrix{\noalign{\vskip5mm}
      &              &{\scriptstyle #5}&              &     \cr
      & #1           & \llrmap           & #2           &     \cr
{\scriptstyle #8}   &\updownarrow& &\updownarrow&{\scriptstyle#6} \cr
      & #4           & \llrmap           & #3           &     \cr
      &              &{\scriptstyle#7}&              &     \cr
\noalign{\vskip5mm} 
            }}

\begin{document}
\begin{titlepage} 
\hfill MPI-PhT/99-51

%\hfill hep-th/... \hfill papref.tex

\vskip25mm
\centerline{\bf DUALITY INDUCED REFLECTIONS AND CPT}

\vskip1cm
\centerline{
Heinrich Saller\footnote{\scriptsize 
hns@mppmu.mpg.de}  
}
\centerline{Max-Planck-Institut f\"ur Physik and Astrophysik}
\centerline{Werner-Heisenberg-Institut f\"ur Physik}
\centerline{M\"unchen}
\vskip15mm

\centerline{\bf Abstract}
\vskip5mm
The linear particle-antiparticle  conjugation $\ty C$
and  position space reflection $\ty P$
as well as the antilinear time reflection $\ty T$
are shown to be inducable by the selfduality of 
representations for the
operation groups $\SU(2)$, $\SL(\C^2)$ and $\R$
for spin, Lorentz transformations and time translations resp.
The definition of a colour compatible linear $\ty{CP}$-reflection for
quarks  as
selfduality induced is impossible since triplet
and antitriplet $\SU(3)$-representations are not linearly equivalent.

\end{titlepage}

 \tableofcontents

%\newpage

\section{Reflections}

\subsection{Reflections}

A reflection will be defined to be an involution
 of a 
finite dimensional vector space
$V$
\begin{eq}{l}
V\stackrel R{\llrmap} V,~~R\o R=\id_V\iff R= R^{-1}\cr
\end{eq}i.e. a realization of the  parity group\footnote{\scriptsize
Since the parity group is used as multiplicative group,
I do not use the additive notation $\Z_2=\{0,1\}$.} 
 $\I(2)=\{\pm1\}$
in the $V$-bijections which is linear for a real space and may be
linear or  antilinear for a complex space
\begin{eq}{rl}
R(v+w)=R(v)+R(w),~ R(\al v)&=\left\{\begin{array}{lll}
\al R(v)&\hbox{for }\al\in\R\hbox{ or }\C&\hbox{(linear)}\cr
\ol\al R(v)&\hbox{for }\al\in\C,&\hbox{(antilinear)}\cr\end{array}\right.
\end{eq}An antilinear reflection for a complex 
space $V\cong\C^n$ is a real linear one for its real forms $V \cong\R^{2n}$.

The inversion of the real  numbers
$\al\lrmap-\al$ is the simplest nontrivial linear reflection, the canonical 
conjugation $\al\lrmap\ol\al$ is the simplest nontrivial antilinear one
being a linear one of $\C$ considered as real 2-di\-men\-sio\-nal space
$\C =\R\pl i\R$.

Any (anti)linear isomorphism $\io:V\map W$
of two vector spaces  defines an (anti)linear  reflection of the direct sum
$V\pl W\stackrel{~~\io\pl\io^{-1}}{\llrmap}V\pl W$ which will be denoted 
in short also by $V\stackrel{\io}{\llrmap}W$.

\subsection{Mirrors}

The fixpoints of a linear reflection 
$V^+_R=\{v\mid R(v)=v\}$
, i.e. the elements with even parity, 
in an $n$-di\-men\-sio\-nal space constitute  a vector subspace, 
the mirror for the reflection $R$,
with dimension $0\le m\le  n$ with the  complement
$V^-_R=\{v\mid R(v)=-v\}$, i.e. the elements with odd parity, 
  for the direct decomposition
$V=V_R^+\pl V_R^-$. 
The central 
reflection $R=-\id_V$ has the origin as a $0$-di\-men\-sio\-nal mirror.
Linear reflections are diagonalizable 
$R\cong{\scriptsize\pmatrix{\bl1_m&0\cr0&-\bl1_{n-m}\cr}}$ with
$(m,n-m)$ the signature
characterizing the degeneracy of $\pm1$ in the spectrum of $R$.
And vice versa: Any direct decomposition $V=V^+\pl V^-$
defines two reflections with the mirror either  $V^+$ or $V^-$.

With $(\det R)^2=1$ any linear reflection 
has either a positive  or a negative orientation.
Looking in the 2-di\-men\-sio\-nal bathroom mirror is formalized by the negatively
oriented
3-space reflection $(x,y,z)\lrmap(-x,y,z)$.
The position space $\R^3$ 
reflection $\rvec x\stackrel {-\bl1_3}\llrmap -\rvec x$ 
with negative orientation or
the Minkowski spacetime translation $\R^4$ reflection 
$x\stackrel {-\bl1_4}\llrmap -x$
with positive orientation 
are central reflections with the origins `here` and `here-now' as 
point mirrors. 
A space reflection $(x_0,\rvec x)\stackrel {\ty P}\llrmap(x_0,-\rvec x)$
in Minkowski space 
or a time reflection $(x_0,\rvec x)\stackrel{\ty T}\llrmap(-x_0,\rvec x)$
have both negative orientation with a 1-di\-men\-sio\-nal time 
and 3-di\-men\-sio\-nal position space mirror resp.

\subsection{Reflections in Orthogonal Groups}

A real linear reflection
$R\cong{\scriptsize\pmatrix{\bl1_m&0\cr0&-\bl1_{n-m}\cr}}$ 
can be considered to be an element of 
an orthogonal group
$\O(p,q)$ for any\footnote{\scriptsize
The orthogonal signature $(p,q)$ 
has nothing to do with the reflection signature $(n,m)$.} $(p,q)$
with $p+q=n$.  A positively oriented
reflection, $\det R=1$, is 
is an element even of the  special orthogonal  groups,  $R\in\SO(p,q)$,
$p+q\ge1$.

Orthogonal groups have discrete
(semi)direct factor parity subgroups $\I(2)$ as seen in the simplest
compact and noncompact examples
\begin{eq}{rlll}
\O(2)\ni&{\scriptsize\pmatrix{
\ep\cos\al& \ep\sin\al\cr
-\sin\al& \cos\al\cr}},&\ep\in\I(2)=\{\pm1\},&\al\in[0,2\pi[\cr
\O(1,1)\ni&\ep'{\scriptsize\pmatrix{
\ep\cosh\be&\ep \sinh\be\cr
\sinh\be& \cosh\be\cr}},&\ep,\ep'\in\I(2),&\be\in\R\cr
\end{eq}In general, the classes of a real  orthogonal groups 
with respect to its special normal subgroup constitute
a reflection group
\begin{eq}{l}
\O(p,q)/\SO(p,q)\cong\I(2)
\end{eq}For real odd dimensional  
spaces $V$, e.g. for position space $\R^3$,
one has direct products of the special groups with
the central reflection group, whereas 
for even dimensional  
spaces, e.g. a
Minkowski  space $\R^4$,
there arise semidirect products (denoted by $\sx$) of the special group with
a reflection group which can be generated by any 
negatively oriented reflection 
\begin{eq}{l}
 \O(p,q)\cong
\left\{\begin{array}{ll}
\I(2)\x\SO(p,q),&p+q=1,3,\dots\cr
&\I(2)\cong\{\pm\id_V\}\cr
\I(2)\sx\SO(p,q),&p+q=2,4,\dots\cr
&\I(2)\cong\{R,\id_V\}\hbox{ with }\det R=-1\cr\end{array}\right.
\end{eq}In the semidirect case the product is given
as follows
\begin{eq}{l}
(I,\La)\in\I(2)\sx\SO(p,q)\then
(I_1,\La_1)
(I_2,\La_2)=(I_1\o I_2,\La_1\o I_1\o \La_2\o I_1)
\end{eq}Obviously, in the semidirect  case
the reflection group  $\I(2)$ is not compatible with
the action of the (special) orthogonal group.
\begin{eq}{l}
p+q=2,4,\dots,~\det R=-1\then \com R{\SO(p,q)}\ne\{0\}
\end{eq}E.g. the group $\O(2)$ is nonabelian, or,
 a space reflection and a time reflection of Minkowski space
is not Lorentz group $\SO(1,3)$ compatible.

For noncompact orthogonal groups there is another discrete reflection group:  
The connected  subgroup $G_0$
(unit connection component and Lie algebra exponent)  of a Lie group $G$ is  normal
with a discrete quotient group $G/G_0$.
The connected components of the full orthogonal groups 
are those of the special groups $\O_0(p,q)=\SO_0(p,q)$.
For the compact case they are
the special groups, for the noncompact ones one has two components
\begin{eq}{rl}
&\SO_0(n)=\SO(n)\cr
pq\ge1\then&
\SO(p,q)/\SO_0(p,q)\cong\I(2)
\cr
\end{eq}

Summarizing: 
A compact orthogonal group gives rise to a reflection group $\I(2)$
\begin{eq}{rl}
 \O(n)&\cong
\left\{\begin{array}{ll}
\{\pm\bl1_n\}\x\SO(n),&n=1,3,\dots\cr
\I(2)\sx\SO(n),&n=2,4,\dots\cr\end{array}\right.\cr
\hbox{with }\I(2)&\cong\{R,\bl1_n\},~~\det R=-1\cr
\end{eq}a noncompact one to a reflection Klein group $\I(2)\x\I(2)$
\begin{eq}{rl}
pq\ge1:~~ \O(p,q)&\cong
\left\{\begin{array}{ll}
\{\pm\bl 1_{p+q}\}\x[\I(2)\sx \SO_0(p,q)],&p+q=3,5,\dots\cr
\I(2)\sx[\{\pm\bl 1_{p+q}\}\x \SO_0(p,q)],&p+q=2,4,\dots\cr\end{array}\right.\cr
\hbox{with }\I(2)&\cong\{R,\bl1_n\},~~\det R=-1\cr
\end{eq}

For a noncompact $\O(p,q)$ with $p=1$ the connected subgroup
is the orthochronous group, 
compatible with the order on the vector space $V\cong\R^{1+q}$, 
e.g. for Minkowski spacetime
\begin{eq}{l}
\O(1,3)\cong\I(2)\sx[\I(2)\x\SO_0(1,3)]\cr
\end{eq}where the reflection  Klein group can
be generated by the central reflection
$-\bl1_4$
and a position space  reflection $\ty P$ 
\begin{eq}{rl}
\I(2)\x\I(2)\cong&\{{\ty P},\bl1_4\}\x\{\pm\bl1_4\}
=\{\pm\bl1_4,{\ty P},{\ty T}=-{\ty P}\},~~[\SO_0(1,3),\ty P]\ne\{0\} \cr
&{\ty P}={\scriptsize\pmatrix{1&0\cr 0&-\bl1_3\cr}},~~
{\ty T}=-\bl1_4\o{\ty  P}={\scriptsize\pmatrix{-1&0\cr 0&\bl1_3\cr}}
\end{eq}

Also the connected subgroup $\SO_0(p,q)$ may contain positively oriented
reflections which are called continuous since they can be written
as exponentials $R=e^l$ with an element of the orthogonal
 Lie algebra\footnote{\scriptsize
 $\log G$ denotes the Lie algebra of the Lie group $G$.}, $l\in\log\SO_0(p,q)$. E.g the  
 central reflections $-\bl1_{2n}\in\SO(2n)$  in 
 even dimensional Euclidean   spaces, e.g. in the 
 Euclidean 2-plane. A negatively oriented  reflection $R$ of a space $V$    
can be embedded as a reflection $R\pl S$ 
 with any orientation of a strictly higher dimensional
space $V\pl W$
\begin{eq}{cl}
V\stackrel R\llrmap V,&\det R=-1\cr
V\pl W\stackrel {R\pl S}\llrmap V\pl W,&\det(R\pl S)=-\det S\cr
\end{eq}where, for  compact orthogonal groups on $V$ and $V\pl W$,  
a reflection $R\pl S$ with $\det S=-1$
is a continuous  reflection, i.e. a rotation. There are the familiar
examples\cite{GARD} for $\O(n)\inmap\SO(n+1)$: Two letter noodles in $L$-form, 
lying with opposite helicity on the
kitchen  table,  can be $3$-space
rotated into each other, or,  a left and a right handed glove
are identical up to Euclidean 4-space rotations. 
The embedding of the central position space
reflection into Minkowski spacetime
can go into a positively or negatively oriented reflection which are both
not continuous, i.e. they are in the discrete Klein reflection group
\begin{eq}{l}
-\bl 1_3\inmap{\scriptsize\pmatrix{\pm1&0\cr0&-\bl1_3\cr}},~~
\{{\ty P},-\bl1_4\}
\subnoteq\O(1,3)/\SO_0(1,3)
\end{eq}

\section{Reflections for Spinors}

The doubly connected groups $\SO(3)$ and $\SO_0(1,3)$
can be complex   represented via their simply connected  covering groups 
$\SU(2)$ and\footnote{\scriptsize
Throughout this paper the group $\SL(\C^2)$ is used  as {\it real}
6-di\-men\-sio\-nal Lie group.}
 $\SL(\C^2 )$ resp.
\begin{eq}{l}
\SO(3)\cong\SU(2)/\{\pm\bl 1_2\},~~
\SO_0(1,3)\cong\SL(\C^2 )/\{\pm\bl 1_2\},~~
\end{eq}The reflection group $\{\pm\bl 1_2\}$
for the $\SO(3)$-classes in $\SU(2)$ 
and the $\SO_0(1,3)$-classes in $\SL(\C^2 )$ 
contains the continuous central $\C^2$-reflection $-\bl1_2=e^{i\pi\si_3}\in\SU(2)$.

\subsection{The  Pauli Spinor Reflection}

The fundamental defining $\SU(2)$-re\-pre\-sen\-ta\-tion   
for the rotations acts on Pauli spinors $W\cong\C^2$ 
\begin{eq}{l}
u=e^{i\rvec\al\rvec\si}\in\SU(2)\hbox{ (Pauli matrices $\rvec \si$)}
\end{eq}They have an invariant
antisymmetric bilinear form (spinor `metric')
\begin{eq}{l}
\ep:W\x W\map\C,~~\ep(\psi^A,\psi^B)=\ep^{AB}=-\ep^{BA},~~A,B=1,2 
\end{eq}which defines an isomorphism
with the dual\footnote{\scriptsize
The linear forms $V^T$ of a vector space $V$ define the dual product
$V^T\x V\map\C$ by $\dprod\om v=\om(v)$ and dual bases by
$\dprod{\d e_j}{e^k}=\de_j^k$.
Transposed mappings $f:V\map W$ are denoted by $f^T:W^T\map V^T$ with
$\dprod{f^T(\om)}v=\dprod\om{f(v)}$.} space $W^T\cong\C^2$
is compatible
with the $\SU(2)$-action - 
on the dual space
as dual representation $\d u$
 (inverse transposed)
\begin{eq}{l}
\Diagr WW{W^T}{W^T} u\ep {\d u}\ep,~~\begin{array}{rl}
\psi^A&\lrmap\ep^{AB}\psi^\star_B\cr 
\d u&=u^{-1T}=u^{\star-1}=\ol u=(e^{-i\rvec\al\rvec\si})^T\cr
u&=\ep^{-1}\o\d u\o\ep\cr
-\rvec\si&=\ep^{-1}\o\rvec \si^T\o\ep\end{array}
\end{eq}$\ep$ connects
the two Pauli representations  with
 reflected  transformations of the spin Lie algebra $\log\SU(2)$, i.e. 
it defines a central reflection for
the three compact
rotation parameters $\rvec\al$
\begin{eq}{c}
e^{i\rvec\al\rvec\si}
\stackrel\ep \llrmap (e^{-i\rvec\al\rvec\si})^T\cr
i\rvec\al\rvec\si\in\log\SU(2)\cong\R^3,~~
\rvec\al\stackrel\ep \llrmap-\rvec\al
\end{eq}and will be called the {\it Pauli spinor reflection} 
\begin{eq}{l}
W\stackrel\ep\llrmap W^T,~~\psi^A\lrmap\ep^{AB}\psi^\star_B,~~
[\ep,\SU(2)]=\{0\}
\end{eq}

The mathematical structure of selfduality
as a reflection generating mechnism is  given in the appendix.

\subsection{Reflections $\ty C$ and $\ty P$ for Weyl Spinors}

The two fundamental $\SL(\C^2 )$-re\-pre\-sen\-ta\-tions
 for the Lorentz group
are the  the left and right handed Weyl representation
on $W_L,W_R\cong\C^2$
with the dual representations on the linear forms $W_{L,R}^T$
\begin{eq}{rllrlll}
\hbox{left:}&\hskip-2mm\la&\hskip-3mm
=e^{(i\rvec\al+\rvec\be)\rvec\si}\in\SL(\C^2),&\hskip-3mm
\hbox{right:}&\hskip-2mm\hat\la&\hskip-3mm=\la^{-1\star}&\hskip-3mm
=e^{(i\rvec\al-\rvec\be)\rvec\si}\cr
\hbox{left dual:}
&\hskip-2mm\d\la=\la^{-1T}&\hskip-3mm=[e^{(-i\rvec\al-\rvec\be)\rvec\si}]^T,
&\hskip-3mm
\hbox{right dual:}&\hskip-2mm\la^{T\star}&\hskip-3mm=\ol\la&\hskip-3mm
=[e^{(-i\rvec\al+\rvec\be)\rvec\si}]^T
\cr
\end{eq}

The Weyl representations with dual bases 
in the conventional notations
 with
dotted and undotted indices\footnote{\scriptsize
The usual strange looking crossover 
association of the letters $\ro l^\star$ and
$\ro r^\star$ for right and left handed dual spinors resp.
will be discussed later.}
\begin{eq}{rlrl}
\hbox{left:}&\ro l^A\in W_L\cong\C^2,&
\hbox{right:}&\ro r^{\dot A}\in W_R\cong\C^2\cr
\hbox{left dual:}&\ro r^\star_A\in W_L^T\cong\C^2,&
\hbox{right dual:}&\ro l^\star_{\dot A}\in W_R^T\cong\C^2\cr
\end{eq}are selfdual
with  the $\SL(\C^2 )$-invariant
volume form on $\C^2$,
i.e. the dual isomorphisms are Lorentz compatible  
\begin{eq}{ll}
\Diagr{W_L}{W_L}{W_L^T}{W_L^T}{\la}{\ep_L}{\d\la}{\ep_L},&~~~~
\Diagr{W_R}{W_R}{W_R^T}{W_R^T}{\hat\la}{\ep_R}{\ol\la}{\ep_R}\cr
\end{eq}

For the Lorentz group
the spinor `metric' will prove to be
related to the {\it particle-antiparticle conjugation},
and will be called {\it Weyl spinor  reflection}, denoted by 
$\ty C\in\{\ep_L,\ep_R\}$
\begin{eq}
{ll}
W_L\stackrel{\ty C}\llrmap W_L^T,&\ro l^A\lrmap\ep^{AB}\ro r_B^\star\cr
W_R\stackrel{\ty C}\llrmap W_R^T,&\ro r^{\dot A}\lrmap\ep^{\dot A\dot B}
\ro l_{\dot B}^\star\cr
\end{eq}

There exist isomorphisms $\de$  between left and right handed
Weyl spinors, compatible with the spin group action,
however not with the Lorentz group $\SL(\C^2 )$
\begin{eq}{l}
\Diagr {W_L}{W_L}{W_R}{W_R}{u_L}\de{u_R}\de,~~~ 
\begin{array}{l}
u_{L,R}=e^{i\rvec\al\rvec\si}\in\SU(2)\cr
\ro l^A\lrmap\de^A_{\dot A}\ro r^{\dot A}\end{array}
\end{eq}They connect representations with
a reflected boost transformation, i.e. 
they define a central reflection for
the three noncompact
boost parameters $\rvec\be$
\begin{eq}{c}
e^{(i\rvec\al+\rvec\be)\rvec\si}
\stackrel\de \llrmap e^{(i\rvec\al-\rvec\be)\rvec\si}\cr
\rvec\si\rvec\be\in\log\SL(\C^2 )/\log\SU(2)\cong\R^3,~~
\rvec\be\stackrel\de \llrmap-\rvec\be
\end{eq}These isomorphisms
induce  nontrivial reflections of the Dirac spinors
$\Psi\in W_L\pl W_R\cong\C^4$
\begin{eq}{l} 
\Psi={\scriptsize\pmatrix{\ro l^A\cr\ro r^{\dot A}\cr}}\stackrel{\de}
\llrmap
{\scriptsize\pmatrix{0&\de^A_{\dot B}\cr\de^{\dot A}_B&0\cr}}
{\scriptsize\pmatrix{\ro l^B\cr\ro r^{\dot B}\cr}}=\ga^0\Psi
\end{eq}with the chiral representation of the Dirac matrices
\begin{eq}{l}
\ga^j={\scriptsize\pmatrix{0&\si^j\cr\d\si^j&0\cr}},~~
\si^j=(\bl1_2,\rvec\si),~~
\d \si^j=(\bl1_2,-\rvec\si)
\end{eq}and will be called
{\it Weyl spinor boost reflections}
$\ty P=\de$, later  used for the central {\it  position space reflection}
representation
\begin{eq}{ll}

W_L\stackrel{\ty P}\llrmap W_R,&\ro l^A\lrmap\de^A_{\dot A}\ro r^{\dot A}\cr
W^T_L\stackrel{\ty P}\llrmap W^T_R,&\ro r_A^\star\lrmap\de_A^{\dot A}
\ro l_{\dot A}^\star\cr
\end{eq}

Therewith all four Weyl spinor spaces are
connected to each other by linear  reflections
\begin{eq}{l}
\Diagrlrud {W_L}{W_R}{W_R^T}{W_L^T}
{\ty P}{\ty C}{\ty P}{\ty C},~~
\begin{array}{ll}
[\ty P,\SL(\C^2 )]\ne\{0\},
&[\ty P,\SU(2)]=\{0\}
\cr
[\ty C,\SL(\C^2 )]=\{0\}&\cr\end{array}
\end{eq}

\section{Time Reflection}

The time representations define the antilinear 
reflection
$\ty T$ for
time translation. 
The different duality with respect to  $\SL(\C^2)$ and 
 Lorentz group representations, 
on the one side, and
time representations, on the other side, leads to the 
nontrivial $\ty{C,P,T}$ cooperation.

\subsection{Reflection $\ty T$ of Time Translations}  

The irreducible
time representations, familiar from the quantum mechanical 
harmonic oscillator with time 
action eigenvalue (frequency) $\om$,
with their duals (inverse transposed) are  complex 1-di\-men\-sio\-nal
\begin{eq}{l}
t\mape e^{i\om t}\in\GL(U),~~
t\mape e^{-i\om t}\in\GL(U^T),~~U\cong\C\cong U^T
\end{eq}They are selfdual (equivalent) with an antilinear dual isomorphism
which is the $\U(1)$-conjugation for a dual basis   $\ro u\in U$, 
$\ro u^\star\in U^T$
\begin{eq}{l}
\Diagr UU{U^T}{U^T}{e^{i\om t}}{\star}{e^{-i\om t}}\star,~~~~
\ro u\lrmap \ro u^\star
\end{eq}The 
antilinear isomorphism $\star$
defines a scalar product which gives rise to the quantum mechanical
probability amplitudes (Fock state for the harmonic oscillator)
\begin{eq}{l}
U\x U\map\C,~~\sprod{\ro u}{\ro u}=\dprod{\ro u^\star}{\ro u}=1 
\end{eq}and
defines the {\it time reflection} $\ty T=\star$
for the time translations
\begin{eq}{c}
e^{i\om t}\stackrel{\ty T}\llrmap e^{-i\om t},~~
t\stackrel{\ty T}\llrmap -t\cr
\end{eq}

\subsection{Lorentz Duality versus Time Duality}

As anticipated in the conventional, on first sight 
strange looking dual Weyl spinor  notation, 
e.g. $\ro l\in W_L$ and $\ro l^\star\in W_R^T$,
the Weyl spinor spaces
$W_{L,R}^T$ with the dual left and right handed $\SL(\C^2)$-re\-pre\-sen\-ta\-tions
are not the spaces with the dual time respesentations
as exemplified in the harmonic analysis of the left and right handed components
in a Dirac field
\begin{eq}{l}
\begin{array}{llcl}
\ro l^{A}(x)
&= \int{d^3 q\over (2\pi)^3}&s({q\over m})^A_C&
~{ e^{xiq}\ro u^C(\rvec q)+ e^{-xiq}\ro a^{\star C}
(\rvec q)\over\sqrt2}
 \cr
\ro l^\star_{\dot A}(x)
&= \int{d^3 q\over (2\pi)^3} &s^\star({q\over m})_{\dot A}^C&
{e^{-xiq}\ro u^\star_C(\rvec q)+ e^{xiq}\ro a_C(\rvec q)\over\sqrt2}
\cr
\ro r^{\dot A}(x)
&= \int{d^3 q\over (2\pi)^3}&s^{\star-1}({q\over m})^{\dot A}_C&
~{ e^{xiq}\ro u^C(\rvec q)- e^{-xiq}\ro a^{\star C}(\rvec q)\over\sqrt2}\cr
\ro r^\star_A(x)
&= \int{d^3 q\over (2\pi)^3} &s^{-1}({q\over m})_A^C&
{e^{-xiq}\ro u^\star_C(\rvec q)- e^{xiq}\ro a_C(\rvec q)\over\sqrt2}
\end{array}\cr
s({q\over m})=\sqrt{{q_0+m\over 2m}}
( \bl 1+{\rvec \si\rvec q\over q_0+m}~),~~q=(q_0,\rvec q),
~~q_0=\sqrt{m^2+\rvec q^2}
\end{eq}Here, $s({q\over m})\in\SL(\C^2)$ is the Weyl representation 
of the boost from the rest 
system of the
particle 
to a frame moving with velocity
${\rvec q\over q_0}$ 
 (solution of the Dirac equation), 
$\ro u^C$ and $\ro a_C$ are the creation operators 
for particle and antiparticles
with spin ${1\over2}$
and opposite charge number $\pm1$ and 3rd  spin direction, 
e.g. for electron and positron, 
 $\ro u_C^\star$ and $\ro a^{\star C}$ are the 
corresponding annihilation operators. 

$\star$ denotes the time representation dual
$U\lrmap U^\star$, and $T$
the Lorentz representation dual $W\lrmap W^T$
(with spinor indices up and down), i.e. for the four types of Weyl spinors 
\begin{eq}{ll}
\Diagrlrud {\ro l^A\in W_L}{\ro l^\star_{\dot A}\in W_R^T=W_L^\star}
{\ro r^{\dot A}\in W_R}
{\ro r^\star_A\in W_L^T=W_R^\star} 
{\rm time~ dual}{\rm Lorentz~ dual}
{\rm time~ dual}{\rm Lorentz~ dual}
\end{eq}Time representation 
duality does not coincide with Lorentz group representation duality.

The antilinear time reflection ($\U(1)$-conjugation) $\ty T=\star$
is  compatible with the action of the little group $\SU(2)$, not with the full
Lorentz group
\begin{eq}{l}
\left.\begin{array}{cc}
W_L\stackrel{\ty T}\llrmap W_R^T,&\ro l^A\lrmap\de^{A\dot A}\ro l^\star_{\dot
A}\cr
W_R\stackrel{\ty T}\llrmap W_L^T,&\ro r^{\dot A}\lrmap\de^{\dot AA}\ro r^\star_A\cr
\end{array}\right\},~~[\ty T,\SL(\C^2)]\ne 0,~~[\ty T,\SU(2)]= 0
\end{eq}

\subsection{The Cooperation of $\ty {C,P,T}$ in the Lorentz Group}

It is useful to summarize the action of the linear Weyl spinor
reflections $\ty C$ (particle-antiparticle conjugation) 
and $\ty P$ (position space central reflection)
and the antilinear time reflection $\ty T$ in the two 
types of commuting diagrams
\begin{eq}{ccc}
\Diagrlrud {W_L}{W_R}{W_R^T} {W_L^T}
{\ty P}{\ty C}{\ty P}{\ty C}&\hbox{ with }&
\Diagrlrud {\ro l^A} 
{\de^A_{\dot A}\ro r^{\dot A}}
{\de^A_{\dot A}\ep^{\dot A\dot B}\ro l^\star_{\dot B}}
{\ep^{AB}\ro r^\star_B}
{\ty P}{\ty C}{\ty P}{\ty C}\cr
\Diagrlrud {W_L}{W^T_R}{W_R} {W_L^T}
{\ty T}{\ty C}{\ty T}{\ty C}&\hbox{ with }&
\Diagrlrud {\ro l^A}
{\de^{A\dot B}\ro l^\star_{\dot B}}
{\de^{A\dot B}\ep_{\dot B\dot A} \ro r^{\dot A}}
{\ep^{AB}\ro r^\star_ B}
{\ty T}{\ty C}{\ty T}{\ty C}\cr
\end{eq}with the compatibilities
\begin{eq}{l}
[\ty C,\SL(\C^2)]=\{0\},~~
[\ty P\hbox{ and }\ty T,\SL(\C^2)]\ne\{0\},~~
[\ty P\hbox{ and }\ty T,\SU(2)]=\{0\}\cr
[\ty C,\ty P]=0,~~
[\ty C,\ty T]=0,~~
[\ty P,\ty T]=0\cr
\end{eq}

The product $\ty{CPT}$ is an antilinear reflection of each Weyl spinor space,
e.g. for the left handed spinors 
\begin{eq}{l}
W_L\stackrel{\ty{CPT}}\llrmap W_L,~~ 
\ro l^A\lrmap \de_{\dot A}^A\ep^{\dot A\dot B}\de_{\dot B B}\ro l^B
\end{eq}involving  an element of the group $\SL(\C^2)$, even of $\SU(2)$
\begin{eq}{l}
\ty{CPT}\sim \de_{\dot A}^A\ep^{\dot A\dot B}\de_{\dot B B}=u_B^A
\cong {\scriptsize\pmatrix{0&1\cr -1&0\cr}}= e^{i{\pi\over2}\si_2}
\in\SU(2)\subnoteq \SL(\C^2)
\end{eq}This element gives - in the used basis - 
for the Lorentz group a $\pi$-rotation 
around the 2nd axis in  position space, i.e. a continuous 
reflection
\begin{eq}{l}
\SU(2)\ni e^{i{\pi\over2}\si_2}\mape {\scriptsize\pmatrix{
-1&0&0\cr 0&1&0\cr 0&0&-1\cr}}\in\SO(3),~~(x,y,z)\lrmap(-x,y,-z)
\end{eq}

The fact that the antilinear $\ty{CPT}$-reflection is 
- up to a number conjugation (indicated by overlining) -
an element of $\SL(\C^2)$, covering the connected Lorentz group $\SO_0(1,3)$,
is decisive for the proof of the well known $\ty{CPT}$-theorem\cite{PAUL, LUD}
\begin{eq}{l}
\ol{\ty{CPT}}\in\SL(\C^2)
\end{eq}

\section{Spinor Induced Reflections}

The linear  spinor reflections $\ep$ for Pauli spinors and
$\ty{C,P}$ for Weyl spinors are inducable on all
irreducible finite dimensional representations of 
$\SU(2)$ and $\SL(\C^2)$ with 
their adjoint groups  $\SO(3)$  and $\SO_0(1,3)$ resp.
via the general procedure:
Given the group $G$ action on two vector spaces its 
tensor product representation 
reads
\begin{eq}{l}
G\x(V_1\ox V_2)\map V_1\ox V_2,~~g\m(v_1\ox v_2)=(g\m v_1)\ox (g\m v_2) 
\end{eq}A realization of the
simple  reflection group $\I(2)=\{\pm1\}$ is either faithful
or trivial.

\subsection{Spinor Induced Reflection of Position Space}

The  reflection $W\stackrel\ep\llrmap W^T$ for
a Pauli spinor space $W\cong\C^2$
induces the central reflection of position
space whose elements come   - in the Pauli representation of position space -
as traceless
hermitian $(2\x 2)$-matrices
\begin{eq}{l}
\rvec x: W\map W,~~\tr \rvec x=0,~~
\rvec x=\rvec x^\star={\scriptsize\pmatrix{
x_3&x_1-ix_2\cr x_1+ix_2&-x_3}}
\end{eq}i.e. as elements\footnote{\scriptsize
The linear mappings $\{V\map W\}$ for finite dimensional vector
spaces are naturally isomorphic to the tensor product
$W\ox V^T$ with the linear $V$-forms $V^T$.}
 of the tensor product $W\ox W^T$ with the
induced $\ep$-reflection 
\begin{eq}{l}
-\rvec\si=\ep^{-1}\o\rvec \si^T\o\ep\then
\rvec x\stackrel\ep\llrmap \ep^{-1}\o\rvec x^T\o\ep=-\rvec  x 
\end{eq}

In the Cartan representation the Minkowski spacetime translations are 
 hermitian mappings from right handed  to left handed spinors
\begin{eq}{l}
x: W_R\map W_L,~~
x=x^\star=
{\scriptsize\pmatrix{
x_0+x_3&x_1-ix_2\cr x_1+ix_2&x_0-x_3}}
\end{eq}i.e.  tensors in 
the product $W_L\ox W_R^T$. The linear $\ty{CP}$-reflection
for Weyl spinors
\begin{eq}{l}
W_L\stackrel{\ty{CP}}\llrmap W_R^T,~~
W_R\stackrel{\ty{CP}}\llrmap W_L^T
\end{eq}induces the position space reflection of Minkowski spacetime
\begin{eq}{l}
\si^j=(\bl 1_2,\rvec\si),~~
\ep^{-1}\o (\si^j)^T\o\ep=\si_j= (\bl 1_2-\rvec\si)\cr
x\cong(x_0,\rvec x)\stackrel{\ty{CP}}\llrmap \ep^{-1}\o x^T\o\ep= 
{\scriptsize\pmatrix{
x_0-x_3&-x_1+ix_2\cr -x_1-ix_2&x_0+x_3}}\cong(x_0,-\rvec x)
\end{eq}

\subsection{Induced Reflections of  Spin Representation Spaces}

All irreducible complex representations of the spin group $\SU(2)$
 with $2J=0,1,2,\dots$ have an invariant bilinear form
arising as a symmetric tensor product of the antisymmetric 
spinor `metric' $\ep$. The bilinear form is
given for the irreducible representation
$[2J]\cong {\OD^{2J}}u$ on the
vector space ${\OD^{2J}}W\cong\C^{2J+1}$
by the corresponding totally symmetric\footnote
{\scriptsize
$\OD$ and $\AND$ denotes symmetrized
and antisymmetrized tensor products.} power
and is antisymmetric for halfinteger spin
and symmetric for integer spin 
\begin{eq}{l}
\ep^{2J}={\OD^{2J}}\ep,~~
\ep^{2J}(v,w)=\left\{\begin{array}{ll}
+\ep^{2J}(w,v),&2J=0,2,4\dots\cr
-\ep^{2J}(w,v),&2J=1,3,\dots\cr\end{array}\right.
\end{eq}

The complex representation spaces for integer spin 
$J=0,1,\dots$, acted upon
faithfully only with the special rotations $\SO(3)\cong\SU(2)/\{\pm\bl1_2\}$,
are direct sums 
of two irreducible real $\SO(3)$-re\-pre\-sen\-ta\-tion spaces $\R^{2J+1}$
where the invariant bilinear form is symmetric and definite,
e.g. the negative definite Killing
form $-\bl1_3$ for the adjoint representation $[2]\cong u\od u$
on $\R^3$. 

The  Pauli spinor reflection 
induces the reflections for the irreducible spin representation spaces  
\begin{eq}{l}
V\cong{\OD^{2J}}W\cong\C^{2J+1}:~~
V\stackrel{\ep^{2J}}\llrmap V^T\cr
\end{eq}

For integer spin (odd dimensional
representation spaces) the two 
real subspaces with irreducible real  $\SO(3)$-re\-pre\-sen\-ta\-tion
come with a trivial $-\bl1_3\mape\bl1_{2J+1}$ and a faithful 
$-\bl1_3\mape-\bl1_{2J+1}\in\O(2J+1)/\SO(2J+1)$ representation
of the central position space reflection, as seen in the diagonalization
of the induced  reflection 
\begin{eq}{l}
{\scriptsize\pmatrix{
0&\ep^{2J}\cr
[\ep^{2J}]^{-1}&0\cr}}=\left\{
\begin{array}{lll}
{\scriptsize\pmatrix{0&1\cr 1&0\cr}}
&\cong{\scriptsize\pmatrix{1&0\cr 0&-1\cr}},&J=0\cr
\cr
{\scriptsize\pmatrix{0&\ep\cr \ep^{-1}&0\cr}},
&
&J={1\over2}\cr
\cr
{\scriptsize\pmatrix{0&-\bl1_3\cr- \bl1_3&0\cr}}
&\cong {\scriptsize\pmatrix{\bl1_3&0\cr 0&-\bl1_3\cr}},&J=1\cr
\cr
\hbox{etc.}&&\end{array}\right.
\end{eq}

The decomposition  for the integer spin representation
spaces uses symmetric and antisymmetric tensor products as 
illustrated for the scalar and vector spin representation
with a Pauli spinor basis
\begin{eq}{cl}
W\stackrel\ep\llrmap W^T,&~~~~\psi^A\lrmap\ep^{AB}\psi_B^\star
,~~J={1\over2} \cr
W^T\ox W\stackrel\ep\llrmap W\ox W^T,&\left\{\begin{array}{rccl}
\psi^\star_A\ox\psi^A &\lrmap&\psi^A\ox\psi^\star_A,&J=0 \cr
\rvec \si^A_B\psi^\star_A\ox\psi^B &\lrmap&
-\rvec \si^A_B\psi^B\ox\psi^\star_A,&J=1 \cr\end{array}\right.
\end{eq}Writing for the tensor (anti)commutator
$[a,b]_\ep= a\ox b+\ep b\ox a$ with $\ep=\pm1$
one has in both cases one trivial and one faithful 
reflection representation 
\begin{eq}{cccl}
[\psi^\star_A,\psi^A]_\ep &\lrmap&\ep[\psi^\star_A,\psi^A]_\ep&J=0 \cr
[\psi^\star_A\rvec \si^A_B,\psi^B]_\ep &\lrmap&
-\ep[\psi^\star_A\rvec \si^A_B,\psi^B]_\ep 
,&J=1 \cr
\end{eq}

\subsection{Induced Reflections of 
Lorentz Group Representation Spaces}

The generating structure of the two Weyl representations
induces  $\ty {C,P}$-reflections of 
$\SL(\C^2)$-re\-pre\-sen\-ta\-tions spaces.

The complex finite dimensional irreducible representations
of the  group  $\SL(\C^2 )$
are characterized by two spins $[2L|2R]$ with integer and halfinteger
$L,R=0,{1\over2},1,\dots$. They are equivalent to the totally symmetric 
products of the left and right handed Weyl representations
\begin{eq}{l}
\hbox{Weyl left: }[1|0]=\la=e^{(i\rvec\al+\rvec\be)\rvec\si},~~
\hbox{Weyl right: }
[0|1]=\hat\la=e^{(i\rvec\al-\rvec\be)\rvec\si}\cr
[2L|2R]\cong{\OD^{2L}}\la\ox{\OD^{2R}}\hat\la\hbox{ acting on
}V\cong{\OD^{2L}}W_L\ox{\OD^{2R}}W_R\cong\C^{(2L+1)(2R+1)}\cr
\end{eq}$[2L|2R]$ and $[2R|2L]$ are equivalent with respect to the
subgroup $\SU(2)$-re\-pre\-sen\-ta\-tions.
The induced reflections are given by the correponding
products of the Weyl spinor reflections.

The real representation spaces for the Lorentz group $\SO_0(1,3)$
are characterized by integer spin
\begin{eq}{l}
L+R=0,1,2,\dots 
\end{eq}They are all generated  by the 
Minkowski representation $[1|1]\cong\la\ox\ol\la$ 
where the complex 4-di\-men\-sio\-nal  representation
space is decomposable into two real 4-di\-men\-sio\-nal ones,
a hermitian and an antihermitian tensor 
\begin{eq}{l}
\C^4\cong W_L\ox W_R^T\ni \ro l\ox\ro l^\star= z= x+i\al\in\R^4\pl i\R^4 
\end{eq}

With Weyl spinor bases
the induced linear reflections for the Minkowski representation  look as follows
(with $\si^j=(\bl1_2,\rvec\si)=\d\si_j$
and $\si_j=(\bl1_2,-\rvec\si)=\d\si^j$)
\begin{eq}{ccc}

\si^j\stackrel{\ty{P}}\llrmap\d\si_j^T,&

\ro l^\star \si^j\ro l\stackrel{\ty{P}}\llrmap
\ro r^\star \d\si_j^T\ro r&\cr

\si_j\stackrel{\ty{C}}\llrmap\d\si_j^T,&
\ro l^\star \si_j\ro l\stackrel{\ty{C}}\llrmap
\ro r \d\si_j^T\ro r^\star&\cr

\si^j\stackrel{\ty{CP}}\llrmap\si_j^T,&
\ro l^\star \si^j\ro l\stackrel{\ty{CP}}\llrmap
\ro l \si_j^T\ro l^\star,&
\ro r^\star \d\si^j\ro r\stackrel{\ty{CP}}\llrmap
\ro r \d \si_j^T\ro r^\star\cr

\end{eq}and can be arranged in combinations of
definite parity, e.g. for $\ty P$
with Dirac spinors in a vector $\ol\Psi\ga^j\Psi$
and an axial vector $\ol\Psi\ga^j\ga_5\Psi$.
The antilinear time reflection has to change in addition the order
in the product 
\begin{eq}{ccc}
\si^j\stackrel{\ty{T}}\llrmap\si_j,&
\ro l^\star \si^j\ro l\stackrel{\ty{T}}\llrmap
\ro l^\star \si_j\ro l,&
\ro r^\star \d\si^j\ro r\stackrel{\ty{T}}\llrmap
\ro r^\star \d \si_j\ro r\cr
\end{eq}

\subsection{Reflections of Spacetime Fields} 

A field $\Phi$ is a mapping from position space $\R^3$
or, as relativistic field, from Minkowski spacetime
$\R^4$ with values in a complex vector space $V$
with the action of a group $G$ both on
space(time) and on $V$. This defines
the action of the group on the field 
$\Phi\mape g\m\Phi={_g\Phi}$ by the commutativity of the diagram
\begin{eq}{l}
\Diagr{\R^3,\R^4}{\R^3,\R^4}VV {O(g)}{_g\bl\Phi}{D(g)}{\bl\Phi}
,~~~~~~\begin{array}{l}
_g\bl\Phi(x)=D(g)\bl\Phi(O(g^{-1}).x)\cr
 \hbox{for }g \in G\end{array}
\end{eq}For position space the external  action group is 
the Euclidean group $\O(3)\sx\R^3$,
for Minkowski spacetime the Poincar\'e group 
$\O(1,3)\sx\R^4$. The value space may have additional interal action groups,
e.g. $\U(1)$, $\SU(2)$ and $\SU(3)$ hypercharge, isospin and colour 
resp. in the
standard model for quark and lepton fields. 

For Pauli spinor fields 
on position space
the
$\O(3)$-action has a direct $\SU(2)$-factor and a
reflection factor $\I(2)$
\begin{eq}{l}
\psi:\R^3\map W\cong\C^2,~~\left\{
\begin{array}{ll}
_u\psi(\rvec x)=D(u)\psi(O(u^{-1}).\rvec x),&u\in\SU(2),~O(u)\in\SO(3)\cr
\psi^A(\rvec x)\stackrel\ep\llrmap\ep^{AB}\psi_B^\star(-\rvec x),&
\hbox{position reflection }\I(2)\end{array}\right.
\end{eq}

Spacetime fields
have the Lorentz group behaviour
\begin{eq}{l}
_\la\Phi(x)=D(\la).\Phi(O(\la^{-1}).x),~~\la\in\SL(\C^2),~~
O(\la)\in\SO_0(1,3)
\end{eq}The antilinear time reflection uses the conjugation
to the time dual field
\begin{eq}{l}
\Phi(x_0,\rvec x)\stackrel{\ty T}\llrmap \Phi^\star(-x_0,\rvec x)
\end{eq}

The  reflections for  Weyl spinor fields on Minkowski spacetime
are
\begin{eq}{crccl}
\ro l^A&(x_0,\rvec x)&\stackrel{\ty P}\llrmap& 
\de^A_{\dot A}\ro r^{\dot A}&(x_0,-\rvec x)\cr
(\ro l^A,\ro r^{\dot A})&(x_0,\rvec x)&\stackrel{\ty C}\llrmap& 
(\ep^{AB}\ro r^\star_B,\ep^{\dot A\dot B}\ro l^\star_{\dot B})
&(x_0,\rvec x)\cr
(\ro l^A,\ro r^{\dot A})&(x_0,\rvec x)&\stackrel{\ty {CP}}\llrmap& 
(\de^A_{\dot A}\ep^{\dot A\dot B}\ro l^\star_{\dot B},
\de_A^{\dot A}\ep^{A B}\ro r^\star_B)&(x_0,-\rvec x)\cr
(\ro l^A,\ro r^{\dot A})&(x_0,\rvec x)&\stackrel{\ty {T}}\llrmap& 
(\de^{A\dot A}\ro l^\star_{\dot A},
\de^{\dot A A}\ro r^\star_A)&(-x_0,\rvec x)\cr
\end{eq}which is inducable on product representations.

\section{The Standard Model Breakdown of $\ty P$ and $\ty{CP}$}

A relativistic dynamics, 
characterized by a Lagrangian for the fields involved, 
may be invariant with respect to an operation group $G$, e.g. the 
$\ty{C,~P}$ and $\ty{ T}$ reflections, or not.
A breakdown of the symmetry can occur 
in two different ways: Either the symmetry is
represented on the field value space $V$, but
the Lagrangian is not $G$-invariant, or there does
not even exist a $G$-re\-pre\-sen\-ta\-tion on $V$. Both cases
occur in the standard model for quark and lepton fields.

\subsection{Standard Model Breakdown of $\ty P$}

The charge $\U(1)$
vertex in electrodynamics for a Dirac electron-positron
field $\Psi$ interacting with an electromagnetic gauge  field $\Ga_j$ 
\begin{eq}{l}
-\Ga_j\ol\Psi\ga^j\Psi=-\Ga_j(\ro l^\star\si^j\ro l+\ro r^\star\d\si^j\ro r)
\end{eq}is invariant under $\ty P$ and $\ty T$ if the
fields have the Weyl spinor induced behaviour given above.

In the standard model of leptons\cite{WEIN} with a left handed  isospin 
doublet field  $\ro L$ 
and a right handed isospin
singlet field $\ro r$
the hypercharge $\U(1)$ and isospin $\SU(2)$  vertex
with gauge fields $A_j$ and $\rvec B_j$ resp. and internal Pauli matrices
$\rvec \tau$
 reads
\begin{eq}{l}
-A_j(\ro L^\star\si^j{\bl 1_2\over2}\ro L
+\ro r^\star\d\si^j \ro r)
+\rvec B_j \ro L^\star\si^j{\rvec\tau\over2} \ro L
\end{eq}All gauge fields are assumed with the
spinor induced reflection behaviour.
The  $\ty P$-invariance
is broken in two different ways: One component of the 
lepton isodoublet, e.g.  $\ro l={1-\tau_3\over 2}\ro L\in W_L^-\cong\C^2$,
can be used together with the right handed isosinglet $\ro r$
as a basis of a Dirac space $\Psi\in W^-_L\pl W_R\cong \C^4$
with a representation of $\ty P$. This is impossible
for the remaining unpaired left handed field  
${1+\tau_3\over 2}\ro L\in W_L^+\cong\C^2$
- here
$\ty P$ cannot even be defined. However, also for the 
left-right pair $(\ro l,\ro r)$  the resulting gauge vertex 
breaks position space reflection $\ty P$  invariance via the familiar 
neutral weak interactions, induced by a vector field $Z_j$
arising in addition to the $\U(1)$-electromagnetic gauge field $\Ga_j$  
\begin{eq}{rl}
-{A_j+B^3_j\over2}\ro l^\star\si^j\ro l
-A_j\ro r^\star\d\si^j \ro r&=
-\Ga_j\ol\Psi\ga^j\Psi
-Z_j\ol\Psi\ga^j\ga_5\Psi
\cr
\hbox{with}&
{\scriptsize \pmatrix{\Ga_j\cr Z_j\cr}}=
{1\over4}{\scriptsize \pmatrix{3&1\cr
-1&1\cr}}
{\scriptsize \pmatrix{A_j\cr B^3_j\cr}}
\end{eq}

There is no parameter involved whose vanishing would lead to a $\ty
P$-invariant dynamics.

\subsection{$\ty{GP}$-Invariance in the  Standard Model of Leptons}

The $\ty {CP}$-reflection
induced by  the spinor `metric'
\begin{eq}{ll}
W_L\stackrel{\ty {CP}}\llrmap W_R^T,&
\ro l^{ A}\lrmap\de^A_{\dot A}\ep^{\dot A\dot B}
\ro l_{\dot B}^\star\cr
W_R\stackrel{\ty {CP}}\llrmap W_L^T,&
\ro r^{\dot  A}\lrmap\de_A^{\dot A}\ep^{ A B}
\ro r_{ B}^\star\cr
\end{eq}has to include also
a linear reflection of  internal operation representations spaces
in the case of Weyl spinors with nonabelian internal 
degrees of freedom.

For isospin $\SU(2)$-doublets  this reflection is given
by the Pauli isospinor reflection discussed above
and is denoted as internal reflection  by $\ty I=\ep$
\begin{eq}{l}
\Diagr UU{U^T}{U^T} u{\ep} {\d u}{\ep},~~
\begin{array}{l}
u\in\SU(2)\hbox{ (isospin)}\cr
\psi^a\stackrel{\ty I}\lrmap\ep^{ab}\psi^\star_b,~~a,b=1,2\cr
-\rvec \tau=\ep^{-1}\o\rvec\tau^T\o\ep\cr
\end{array} 
\end{eq}Therewith the linear $\ty{GP}$-reflection
as particle-antiparticle conjugation including nontrivial isopsin eigenvalues 
\begin{eq}{l}
\ty G=\ty{IC},~~
\ty {GP}=\ty{ICP}
\end{eq}reads for left handed Weyl
spinors isospinors 
\begin{eq}{l}
W_L\ox U\stackrel{\ty {GP}}\llrmap W_R^T\ox U^T,~~
\ro L^{ Aa}\lrmap\de^A_{\dot A}\ep^{\dot A\dot B}\ep^{ab}
\ro L_{\dot B b}^\star\cr
\end{eq}The antilinear $\ty T$-reflection uses the $\U(2)$-scalar product
\begin{eq}{l}
U\stackrel\star\llrmap U^T,~~U\x U\map\C,~~\sprod{\psi^a}{\psi^b}=\de^{ab}\cr
W_L\ox U\stackrel{\ty {T}}\llrmap W_R^T\ox U^T,~~
\ro L^{ Aa}\lrmap\de^{A\dot B}\de^{ab}
\ro L_{\dot B b}^\star\cr
\end{eq}The isospin dual coincides with the time dual $U^T=U^\star$.

In the product $\ty{CPT}$ there arises - in the basis chosen -
an isospin transformation 
$\ep^{ac}\de_{cb}\cong e^{i{\pi\over2}\tau_2}\in\SU(2)$
\begin{eq}{l}
W_L\ox U\stackrel{\ty{ICPT}}\llrmap W_L\ox U,~~
\ro L^{ Aa}\lrmap\de^{A\dot B}\de_{\dot BB}\ep^{ac}\de_{cb}\ro L^{Bb}
\end{eq}decisive to prove the $\ty{GPT}$-theorem with
\begin{eq}{l}
\ol{\ty{ICPT}}\in \SU(2)\x\SL(\C^2) 
\end{eq}

With the spinor induced reflection behaviour for the gauge fields
the standard model for leptons, i.e. with internal hupercharge-isospin action, 
allows the representation of $\ty{GP}$ and $\ty T$ 
with the gauge vertex above being $\ty{GP}$ and $\ty T$ invariant.

\subsection{$\ty{CP}$-Problems for Quarks}

If quark triplets and antitriplets which come
with the dual defining $\SU(3)$-re\-pre\-sen\-ta\-tions, are included in
the standard model, an extended $\ty{CP}$-reflection has to 
employ a linear reflection $\ga$ between dual representation spaces
of colour $\SU(3)$, i.e. an $\SU(3)$-invariant bilinear form   
of the representation space
\begin{eq}{l}
\Diagr UU{U^T}{U^T}{D(u)}{\ga} {\d D( u)}{\ga},~~
\begin{array}{l}
D:\SU(3)\map\GL(U)\hbox{ (colour representation)}\cr
\ga^{-1}\o D(u)^T\o\ga=D(u^{-1})
\hbox{ for all }u\in \SU(3)\cr
\end{array} 
\end{eq}

The situation for isospin $\SU(2)$ and colour $\SU(3)$ is completely different
with respect to the existence of such a linear dual isomorphism $\ga$:
All irreducible $\SU(2)$-re\-pre\-sen\-ta\-tions $[2T]$ with isospin
$T=0,{1\over 2},1,\dots$ have an - up to a scalar factor -  unique invariant 
bilinear form ${\OD^{2T}}\ep$ as product of the spinor `metric', discussed
above. 

That is not the case for the colour representations.
Some representations are linearly selfdual, some are not.

The complex  irreducible representations
of $\SU(3)$
are characterized by $[N_1,N_2]$ with  two integers 
$N_{1,2}=0,1,2,\dots$. They arise from 
the two fundamental triplet representations, dual to each other
and 
parametrizable with eight Gell-Mann matrices $\rvec\la$
\begin{eq}{l}
\hbox{triplet: }[1,0]=u=e^{i\rvec\ga\rvec \la},~~
\hbox{antitriplet: }
[0,1]=\d u=u^{-1T}=(e^{-i\rvec\ga\rvec \la})^T\cr
[N_1,N_2]\hbox{ acting on vector space }U\hbox{ with }
\dim_\C U={(N_1+1)(N_2+1)(N_1+N_2+2)\over2}
\end{eq}Dual representations have reflected integer values
$[N_1,N_2]\lrmap[N_2,N_1]$.
Only those $\SU(3)$-re\-pre\-sen\-ta\-tions whose weight diagram
is central reflection symmetric in the real 2-dimensional
weight vector space (appendix)
have one, and only one, $\SU(3)$-invariant
bilinear form\cite{LIE8}, i.e. they are linearly selfdual. 
Dual representations have weights which
are reflected to each other
\begin{eq}{rl}
\weights[N_1,N_2]\stackrel{-\bl1_2}\llrmap
\weights[N_2,N_1]\cr
\end{eq}Therefore, one obtains as
 selfdual irreducible $\SU(3)$-re\-pre\-sen\-ta\-tions
\begin{eq}{rl}
\weights[N_1,N_2]=-\weights[N_1,N_2]\iff &N_1=N_2=N\cr
\then&\dim_\C U=(N+1)^3=1,8,27,\dots
\end{eq}E.g. for the octet $[1,1]$ as adjoint $\SU(3)$-re\-pre\-sen\-ta\-tion,
the Killing form defines its selfduality. 

A general remark (appendix): The Lie group $\SL(\C^{r+1})$ with its 
maximal compact subgroup $\SU(r+1)$ of rank $r$
is defined as invariance group
of the $\C^{r+1}$-volume elements which are totally antisymmetric
$(r+1)$-linear forms
$\ep^{a_1\dots a_{r+1}}$. Their complex finite dimensional irreducible
representations are characterized by $r$ integers
$[N_1,\dots,N_r]$ with the dual representations
having the reflected order $[N_r,\dots,N_1]$.
The weights (eigenvalues) for dual representations
are related to each other by the central weight space reflection $-\bl1_r$
which defines the linear  particle-antiparticle conjugation $\ty I$
for $\SU(n)$. 
Only for $n=2$ (isospin $\SU(2)$)
all representations $[N=2T]$ are selfdual with their invariant bilinear
form arising from $\ep^{ab}$
for $[1]$. The $n=2$ selfduality of the doublet
$u(2)\cong\d u(2)$ is replaced for $n=3$ 
by the equivalence of antisymmetric triplet
square and antitriplet representation   $u(3)\and u(3)\cong \d u(3)$, 
i.e. $3\and 3\cong \ol 3$,
with the obvious generalization
${\AND^{r}}u(r+1)\cong\d u(r+1)$ for general rank $r$.

Obviously all $\SU(r+1)$-representations have an invariant sesquilinear form,
the $\SU(r+1)$ scalar product. However, this
antilinear structure cannot define a linear 
particle-antiparticle conjugation.

It is impossible to define a $\ty{CP}$-extending 
 duality induced linear  $\ty{GP}$-reflection for the irreducible complex 3-dimensional 
quark representation spaces
since there does not exist a colour $\SU(3)$-invariant
bilinear form of the triplet space $U\cong\C^3$. Or equivalently:
There does not exist a $(3\x 3)$-matrix $\ga$ for
the reflection $-\rvec\la=\ga^{-1}\o\rvec\la^T\o\ga$
of all eight Gell-Mann matrices.
Therewith there arise also problems to define an $\SU(3)$-compatible
time reflection for quark triplet fields.
Could all that be the reason for the
breakdown of $\ty{CP}$-invariance coming on the quark field sector
and its parametrization (e.g. Cabibbo-Kobayashi-Maskawa) with three families of
colour triplets?

\appendix
\section{Central Reflections of Lie Algebras}

A representation  
of a group $G$ on a vector space $V$ is {\it selfdual}
if it is equivalent to its dual representation,
defined by the inversed transposed action 
on the linear forms $V^T$
\begin{eq}{l}
\left.\begin{array}{l}
D:G\map\GL(V)\cr
\d D:G\map\GL(V^T)\end{array}\right\}
,~~\d D(g)= D(g^{-1})^T
\end{eq}i.e. if the following diagram
with a linear or antilinear isomorphism $\ze:V\map V^T$ commutes
with the action of all group elements
\begin{eq}{l}
\Diagr VV{V^T}{V^T}{D(g)}{\ze}{\d D(g)}{\ze},~~
~~~\ze^{-1}\o D(g)^T\o\ze=D(g^{-1})
\hbox{ for all }g\in G\cr
\end{eq}Selfduality is equivalent to the existence of a nondegenerate
bilinear (for linear $\ze$) or sesquilinear form (for antilinear $\ze$) of
the vector space $V$ 
\begin{eq}{rl}
V\x V\map\C,&\ze(w,v)=\dprod{\ze(w)}v\cr
\hbox{selfdual}&\ze(g\m w,g\m v)=\ze(w,v),~~g\m v=D(g).v
\end{eq}

For the Lie algebra $L=\log G$ of a Lie group $G$ 
with dual representations 
in the endomorphism algebras $\AL(V)$ and $\AL(V^T)$
which are negative transposed to each other
\begin{eq}{l}
\left.\begin{array}{l}
\cl D: L\map\AL(V)\cr
\d{\cl D}:L\map\AL(V^T)\cr\end{array}\right\},
~~\d{\cl D}(l)=-\cl  D(l)^T\cr
\end{eq}a selfduality isomorphism, 
i.e. the reflection $V\stackrel\ze\llrmap V^T$
fulfills
\begin{eq}{l}
\ze(l\m w,v)=-\ze(w,l\m v),~~l\m v=\cl D(l).v
\end{eq}and defines the {\it  central reflection of the Lie algebra}
in the representation 
\begin{eq}{l}
\Diagr VV{V^T}{V^T}{\cl D(l)}{\ze}{\d{\cl D}(l)}{\ze},~~
\ze^{-1}\o \cl D(l)^T\o\ze=-\cl D(l)
\hbox{ for all }l\in \log G\cr
\end{eq}

With Schur's lemma, an irreducible complex finite dimensional
representation of a group or Lie algebra can have at most
- up to a constant - one invariant bilinear and one invariant
sesquilinear form. E.g. Pauli spinors for $\SU(2)$ have both,
$\ep^{AB}$ (bilinear) and $\de^{AB}$ (sesquilinear, scalar product), 
$A,B=1,2$, quark triplets have only a scalar product
$\de^{ab}$, $a,b=1,2,3$, Weyl spinors for $\SL(\C^2)$ have only
the bilinear `metric' $\ep^{AB}$.

For a simple Lie algebra $L$  of rank $r$, the weights
(eigenvalue vectors for
a Cartan subalgebra)  of dual representations $\cl D$ and $\d{\cl D}$   
are related to each other by the central reflection of the weight vector space $\R^r$
\begin{eq}{l}
\weights \cl D[L]\stackrel{-\bl1_r}\llrmap \weights \d{\cl D}[L]
\end{eq}which may be induced by a linear isomorphism
$\ze$ of the dual representation spaces.
Therewith: Such a linear  isomorphism for an $L$-representation
exists\cite{LIE8} if, and only if,
the weights of the representation $\cl D:L\map\AL(V)$ are invariant 
under central reflection
\begin{eq}{l}
V\stackrel\ze\llrmap V^T\iff \weights \cl D[L]=-\weights \cl D[L]
\end{eq}

\end{document}